\documentclass[twocolumn,amsmath,amsfonts,amssymb]{revtex4}
\usepackage{graphicx}
\def\beq{\begin{equation}}
\def\eeq{\end{equation}}
\def\bea{\begin{eqnarray}}
\def\eea{\end{eqnarray}}

\begin{document}

\title{ Phase transitions of hairy black holes in massive gravity and thermodynamic behavior of charged AdS black holes in an extended phase space}

\author{Behrouz Mirza}
\email{b.mirza@cc.iut.ac.ir}
\affiliation{Department of Physics, Isfahan University of Technology, Isfahan, 84156-83111, Iran}
\author{Zeinab Sherkatghanad}
\email{ z.sherkat@ph.iut.ac.ir}
\affiliation{Department of Physics, Isfahan University of Technology, Isfahan, 84156-83111, Iran}

\begin{abstract}
We study the thermodynamic behavior of  static and spherically symmetric hairy black holes in massive gravity. In this case, the black hole is surrounded in a spherical cavity with a fixed temperature on the surface. It is observed that these black holes have a phase transition similar to the liquid-gas phase transition of a Van der Waals fluid. Also, by treating the cosmological constant $\Lambda$ as a thermodynamic pressure $P$, we study the thermodynamic behavior of charged anti-de Sitter black holes in an ensemble with a pressure of $P$ and an electric potential $\Phi$ as the natural variables. A second order phase transition is observed to take place for all the values of the electric potential $\Phi$.
\end{abstract}
\maketitle

%%%%%%%%%%%%%%%%%%%%%%%%%%%%%%%%%%%%%%%%%%%%%%%%%%%%%%%%%%%%%%%%%%%

%%%%%%%%%%%%%%%%%%%%%%%%%%%%%%%%%%%%%%%%%%%%%%%%%%%%%
\section{Introduction}
%%%%%%%%%%%%%%%%%%%%%%%%%%%%%%%%%%%%%%%%%%%%%%%%%%%%%
Black holes have come to be known over the past decades to be, indeed, thermodynamic objects that can be described by a physical temperature and an entropy \cite{TE}. The first attempt in order to explore the thermodynamic properties \cite{w, p} and instability of anti-de Sitter (AdS) black holes date back to the paper written by Hawking and Page \cite{HP}. Based on this successful idea, there have been several works addressing the critical behavior and  nature of phase transitions related to  black holes \cite{G, C1, C2, M1, M2, 2, 3, 4, 5, 6, 7, lala, nicol}.\\
\indent The thermodynamic properties of  static and spherically symmetric hairy black holes  was studied as the solution of massive gravity action in \cite{mass1}. It has been shown that for the negative values of the hair parameter, the phase structure is isomorphic to a Reissner-Nordstrom  (RN) black hole in the canonical ensemble by applying York's procedure \cite{mass2, mass3, mass4,action1,action2}. These properties motivated us to investigate the resemblance between the critical behavior of these black holes and that of  the Van der Waals fluid \cite{Mann1,Mann2, Mann3, Mann4, Mann5, Niu1, Mirza1, zou, cai}. Also, we computed the critical exponents by using two different methods described in \cite{Mann1, Mann2,B1, B2, B3, B4}.\\
\indent Recently, the idea of treating the cosmological constant $\Lambda$ as a thermodynamic pressure in the first law of  black hole thermodynamics has attracted a lot of attention \cite{D1, D2, D3, D4, D5}. Considering an extended phase space thermodynamics seems more meaningful for different reasons. The most important one is the satisfaction of the Smarr formula. Also, including $\Lambda$ as a thermodynamic pressure in the first law leads us to consider the critical behavior of the black hole system by redefining new types of ensembles. In these circumstances, the black hole mass known as an internal energy is replaced by enthalpy. Recently, the critical behavior of charged AdS black holes such as Reissner-Nordstrom-AdS and Kerr-Newman-AdS black holes has been investigated in the extended phase space (while the  cosmological constant $\Lambda$ is included as a thermodynamic variable) and in the canonical ensemble \cite{Mann1, Mann2, GS, GS1}.  The critical behaviors are analogous to the liquid-gas phase transition in the Van der Waals fluid. Also, it has been shown that the critical behavior of RN-AdS and Gauss-Bonnet-AdS black holes at the vicinity of the critical point is similar to the Van der Waals fluid in both the nonextended phase space and the canonical ensemble \cite{Mann3, GS}. Moreover, it is interesting  to study the thermodynamic behavior of these charged AdS black holes in the extended phase space and an ensemble with pressure and electric potential $\Phi$ as the natural variables. \\
\indent The RN-AdS black hole experiences just a second order phase transition, while $\Lambda$ is not considered as a thermodynamic pressure in the grand canonical ensemble \cite{B1}. Here, we investigate the thermodynamic behavior of Gauss-Bonnet-AdS black holes in the grand canonical ensemble. Also, the thermodynamic behavior of Reissner-Nordstrom-AdS in the extended phase space and an ensemble with pressure and electric potential $\Phi$ as the natural variables is studied. The results show that a second order phase transition takes place which is independent of  the values of the electric potential $\Phi$.\\
\indent This paper is organized as follows: In Sec. II, we study the thermodynamic behavior of hairy black holes in massive gravity and  calculate their critical exponents. In Sec. III, we study the  Gibbs free energy $G$ for Reissner-Nordstrom-AdS and Gauss-Bonnet-AdS black holes in the ensemble in which  pressure  and electric potential are considered as  variables.

\section{The critical behavior of hairy black holes in massive gravity}
We start by reviewing the thermodynamic quantities of hairy black holes with spontaneous Lorentz breaking as the solution of massive gravity action. The action in massive gravity is given by \cite{action1,action2}
\bea \label{action}
  I&=&\int _{\cal{M}}  d^4 x \sqrt{g} [-\frac{1}{16 \pi} R+\Lambda^4 {\cal{F}}(X,W_{ij})],\\\nonumber
&-&\int_{\partial M} d^3 x \sqrt{\gamma} \frac{1}{8 \pi} K
\eea
where
\bea
 X&=&\Lambda ^{-4} g^{\mu \nu} \partial _{\mu} \phi ^0 \partial _{\nu} \phi ^0,\\\nonumber
V^i&=&\Lambda ^{-4} g^{\mu \nu} \partial ^{\mu} \phi ^i \partial _{\nu} \phi ^0,\\\nonumber
X&=&\Lambda ^{-4} g^{\mu \nu} \partial _{\mu} \phi ^i \partial _{\nu} \phi ^j -\frac{V^i V^j}{X}.
\eea
Here $\mu$ and $\nu$ specify the spacetime components, and $\phi^{\mu}$ is the four scalar fields. The second integral indicates the Gibbons-Hawking-York boundary term in which $K$ is the trace of the extrinsic curvature.\\
The static spherically symmetric black hole solutions read as follows \cite{mass1, mass2}:
\bea \label{1}
   ds^{2}=\alpha(r)dt^2+\rho(r)dr^2+r^2(d\theta ^2+sin ^2\theta d\varphi ^2),
\eea
where
\bea
   &&\phi^0=\Lambda ^2 [-it+h(r)],\\\nonumber
   &&\phi^i=\phi(r) \frac{\Lambda ^2 x^i}{r},\\\nonumber
    && {\cal{F}}=\frac{12}{\lambda X}+6 (\frac{2}{\lambda}+1) \omega_1-  \omega_1 ^3 +3  \omega_1  \omega_2- 2 \omega_3+12.
\eea
Here $\lambda$ is a positive constant and $\omega_n=Tr(W^n)$. If we consider the above metric and equation of motion of action in Eq. (\ref{action}), the black hole solution is
\bea \label{2}
    &&\alpha(r)=1-\frac{2M}{r}-\frac{Q}{r^\lambda},\\
    &&\rho(r)=\frac{1}{\alpha(r)},\\
    &&h(r)=\pm \int \frac{dr}{\alpha} [1-\alpha  (\frac{Q}{12 m^2} \frac{\lambda (\lambda-1)}{r^{\lambda+2}}+1)^{-1}]^{\frac{1}{2}},\\
    &&\phi(r)=r,
\eea
where $M$ is the ADM mass and $Q$ is the hair parameter. The procedure to obtain the equilibrium thermodynamics needs to enclose the asymptotically flat black hole within a finite volume surface and then send the surface to infinity \cite{mass3, mass4}.\\
The Hawking temperature and entropy are given by
\bea \label{3}
     &&T_H=\frac{\partial_r \alpha}{4\pi}|_{r_+}=\frac{1+(\lambda -1)\frac{Q}{r_+ ^\lambda}}{4\pi r_+},\\
   &&S=4 \pi r_+^2.
\eea
One is able to show that the first law of the black hole system satisfies and we have $dM=TdS+\Phi d|Q|$ for these black holes, where $\Phi=\frac{r_+^{1-\lambda}}{2}$ is the scalar charge potential. Also, the Hawking temperature for an observer at the position $r$ can be expressed by
\bea \label{4}
 T(r)=\frac{T_{H}}{\sqrt{\alpha}}=\frac{1}{4\pi r_{+}}\frac{1+(\lambda -1)\frac{Q}{r_+ ^\lambda}}{\sqrt{1-\frac{r_+}{r}+\frac{Q}{r_+ ^{\lambda -1} r}-\frac{Q}{r^\lambda}}}.
\eea
If we consider a spherical cavity of radius $r_B$ as the boundary and further assume that $Q$ is a conserved quantity in the cavity, we can calculate the above equation at $r=r_{B}$. Thus, the temperature at this boundary $r=r_{B}$ is
\bea \label{5}
    \bar{T}(x,\bar{Q})=\frac{1+(\lambda-1)\frac{\bar{Q}}{x^\lambda}}{x \sqrt{1-x}\sqrt{1+\frac{\bar{Q}}{x^{\lambda -1}}\frac{1-x^{\lambda -1}}{1-x}}},
\eea
where $x=\frac{r_+}{r_B}$, $\bar{Q}=\frac{Q}{r_B ^{\lambda}}$ and $\bar{T}=4\pi r_B T$.  Also, the black hole's scalar charge potential is given by
\bea \label{6}
\bar{\Phi}=\frac{1}{\sqrt{(1-x)(1+\frac{\bar{Q}}{x^{\lambda -1}}\frac{1-x^{\lambda -1}}{1-x})}}  \frac{1-x}{x}.
\eea

\subsection{Equation of state and Gibbs free energy}

Using Eqs. (\ref{5}) and (\ref{6}), we have the possibility to write the equation of state, $\bar{Q}$ as a function of $\bar{\Phi}$ and $\bar{T}$ for $\lambda =2$, in the following form:
\bea \label{7}
\bar{Q}&=&\frac {(1 - \bar{\Phi} \zeta +\zeta ^2 )^2} {1458 \bar{T}^3 \zeta ^4}  \   [-162 \bar{T} \zeta ^2\\\nonumber
&&+\frac {(1 - \bar{\Phi} \zeta +\zeta ^2 )^3 (-1 + \bar{\Phi} \zeta + 3 \bar{T} \zeta-\zeta ^2) } {81 \bar{T} \zeta ^2} \\\nonumber
 \\\nonumber
&&-9  \  (1 - \bar{\Phi} \zeta + \zeta ^2 ) (1 - \bar{\Phi} \zeta - 3 \bar{T} \zeta+\zeta ^2)\\\nonumber
&&(4 + \frac { (1 - \bar{\Phi} \zeta + \zeta ^2)^4} {81 \bar{T}^2 \zeta ^4})^\frac{1}{2}],
\eea
here,
\bea \label{7a}
\zeta=\frac{2^{1/3} \bar{\Phi}^3}{(-2 \bar{\Phi}^6+27 \bar{\Phi}^2 \bar{T}^2+3 \sqrt{3} \sqrt{-4 \bar{\Phi}^8 \bar{T}^2+27 \bar{\Phi}^4 \bar{T}^4})^{1/3}}.
\eea

\begin{figure}
\includegraphics[angle=0,width=8cm,height=7cm]{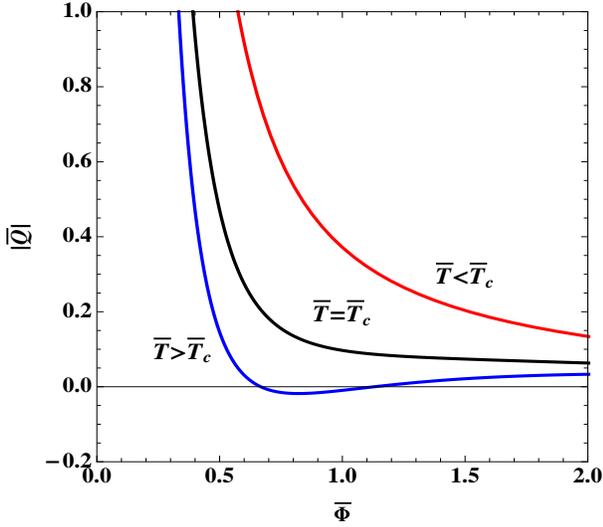}\\
\caption{\it{Scalar charge $\mid\bar{Q}\mid$ with respect to
$\bar{\Phi}$ for $\bar{T}>\bar{T}_{c}$, $\bar{T}=\bar{T}_{c}$, and $\bar{T}<\bar{T}_{c}$.}} \label{figure 1}
\end{figure}

\begin{figure}
\includegraphics[angle=0,width=8cm,height=7cm]{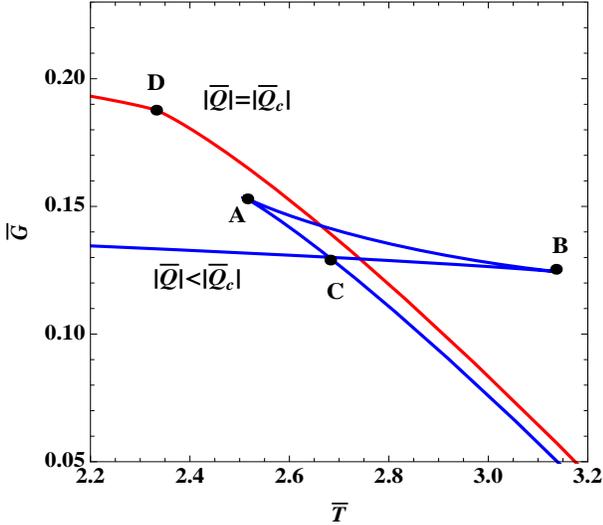}\\
\caption {\it{ Gibbs free energy $\bar{G}$ with respect to $\bar{T}$
for $\mid\bar{Q}\mid<\mid\bar{Q}_{c}\mid$ and $\mid\bar{Q}\mid=\mid\bar{Q}_{c}\mid$. }}\label{figure 2}
\end{figure}

Also, the first law of black hole thermodynamics is described by
 \bea \label{7b}
 d\bar{M}=\bar{T} d \bar{S} + \bar{\Phi} d | \bar{Q}|,
 \eea
where $\bar{M}=2(1-\sqrt{\alpha(r_B)})$ and $\bar{S}=\frac{x^2}{2}$.\\
The scalar charge $\mid\bar{Q}\mid$ is depicted with respect to $\bar{\Phi}$ as $"P-V"$ diagram for $\bar{T}=\bar{T}_{c}$, $\bar{T}>\bar{T}_{c}$, and $\bar{T}<\bar{T}_{c}$ in Fig. \ref{figure 1}. This diagram demonstrates an inflection point for $\bar{T}>\bar{T}_c$ and a critical point at $T=T_c$, similar to the Van der Waals fluid. The critical values of thermodynamic quantities at $T=T_c$ can be calculated by using the two equations  $(\frac{\partial \bar{Q}}{\partial \bar{\Phi}})_c=0$ and $(\frac{\partial ^2 \bar{Q}}{\partial \bar{\Phi}^2})_c=0$.  Also, one is able to calculate the critical values of thermodynamic parameters by calculating the discriminant of the denominator of heat capacity in the canonical ensemble (fixed $Q$) for $\lambda=2$,
\bea \label{8}
 &&\bar{Q}_{c}=\frac{1}{-9-4\sqrt{5}},\\
 &&x_c=5-2\sqrt{5},\\
 &&\bar{T}_c=\frac{2}{5\sqrt{85-38\sqrt{5}}},\\
 &&\bar{\Phi}_c=(\frac{2}{\sqrt{5}}+1)(\sqrt{5-2\sqrt{5}}).
\eea

The Gibbs free energy is defined by
\bea \label{8a}
\bar{G}=1 - \frac { (1 + \frac {\bar{Q}} {x^2}) x}{4 \sqrt {\frac {(1 - x) (\bar{Q} + x)} {x}}}-\sqrt {\frac {(1 - x) (\bar{Q} + x)} {x}},
\eea
where $\bar{G}=\frac{2 G}{r_B}$. Considering Eqs. (\ref{5}) and (\ref{8a}), we can describe the thermodynamic behavior of $\bar{G}$ with respect to $\bar{T}$ in Fig. \ref{figure 2}. At $\mid \bar{Q} \mid<\mid \bar{Q}_{c} \mid$, the black hole experiences just a first order phase transition at point $C$ and the second order phase transitions at points $A$ and $B$ which are in a metastable state. For $\mid \bar{Q} \mid>\mid \bar{Q}_{c} \mid$, the black hole phase transition is removed. On the other hand, at $\mid \bar{Q} \mid=\mid \bar{Q}_{c} \mid$, we have the critical point $D$, which is related to the second order phase transition with the critical values of thermodynamic quantities in Eq. (\ref{8}). Also, we expand our calculations for $\lambda =3, 4,$ and the results are similar to $\lambda =2$. In the next subsection, we calculate the critical exponents at this critical point.\\

\subsection{Critical exponents}

The critical exponent $\alpha$ is associated with the singular behavior of $C_{\bar{\Phi}}$. Since the specific heat at fixed $\bar{\Phi}$, $C_{\bar{\Phi}}$ is not divergent at this critical point, we find the critical exponent $\alpha =0$. If we expand the equation of state $\bar{Q}(\bar{T},\bar{\Phi})$ near the critical values of $\bar{T}_c$ and $\bar{\Phi}_c$, we have
\bea \label{9}
 q=1-5.23 t+9.47 \varphi t- 1.54 \varphi ^3+.....,
\eea
where $q=\frac{\bar{Q}}{\bar{Q}_{c}}$, $t=\frac{\bar{T}}{\bar{T}_c}-1$ and $\varphi=\frac{\bar{\Phi}}{\bar{\Phi}_c }-1$. If the above series is differentiated with respect to $\varphi$ at a  fixed $t$ and considering
Maxwell's equal area law $\oint \bar{\Phi}d\bar{Q}=0$ which is demonstrated in \cite{G}, we have the following equation:
\bea \label{9a}
 0=\int^{\varphi2} _{\varphi1} \varphi (9.47 t- 1.54\times 3 \varphi ^2) d\varphi,
\eea
where $\varphi _1$ and $\varphi_2$ are the electric potentials of the black hole in two different phases. Since  pressure $\bar{Q}$ remains constant during the phase transition, we get
\bea \label{9b}
 q&=&1-5.23 t+9.47 \varphi_1 t- 1.54 \varphi_1 ^3\\\nonumber
 &=&1-5.23 t+9.47 \varphi_2 t- 1.54 \varphi_2 ^3.
\eea
 The nontrivial solution of Eqs. (\ref{9a}) and (\ref{9b}) is given by
\bea \label{10}
 \varphi_1=-\varphi_2= 2.47 \sqrt{t}.
\eea
Based on the critical exponent $\beta$, which is defined by $\eta\propto (\varphi_2 -\varphi _1) \propto t^{\beta}$, we get $\eta \propto \sqrt{t} \Rightarrow \beta=\frac{1}{2}$. For the critical exponent $\gamma$ described by $k_{\bar{T}}\propto t^{-\gamma}$, we have
\bea \label{11a}
(\frac{\partial \bar{\Phi}}{\partial \bar{Q}})_T=\frac{1}{9.47 t}\frac{\bar{\Phi}_c}{\bar{Q}_c},
\eea
so
\bea \label{11}
 k_{\bar{T}}=-\frac{1}{\bar{\Phi}}(\frac{\partial \bar{\Phi}}{\partial \bar{Q}})_T\propto -\frac{1}{9.47 \bar{Q}_c}\frac{1}{t} \ \Rightarrow \ \gamma=1.
\eea
Also, the critical isotherm $\delta$ described at $\bar{T}=\bar{T}_c$ is $q-1=-1.54 \varphi ^3\ \Rightarrow \delta=3$. These critical exponents of the hairy black hole correspond to the Van der Waals fluid.  Now, we are in a position to calculate the critical exponents by using a different method described in \cite{B1, B2, B3} near the critical point. \\
Let us consider the critical exponent $\alpha$. Since the specific heat $C_{\bar{\Phi}}$ is not  divergent at the critical point, we have $\alpha =0$. Consider the parameters $x$ and $\bar{T}$ around the critical point
\bea \label{11a}
 &&x=x_c (1+\Delta),\\
 &&\bar{T}=\bar{T}_c (1+\epsilon),
\eea
here, $\Delta, \epsilon \ll 1$. If  $\bar{\Phi}$ in Eq. (\ref{6}) is expanded around $x_c$, we have
\bea \label{11b}
\bar{\Phi}(x)&=&\bar{\Phi}(x_c)+[(\frac{\partial \bar{\Phi}}{\partial x})_{\bar{Q}}]_{x=x_c} (x-x_c) \\\nonumber
&+& higher\ order\ terms.
\eea
The expansion of $\bar{Q}(x,\bar{T})$ in Eq.(\ref{7}) near the critical values of $x_c$ and $\bar{T}_c$ is given by
\bea \label{11c}
\bar{Q}(x)&=&\bar{Q}(x_c)+[(\frac{\partial \bar{Q}}{\partial x})]_{(x=x_c, \bar{T}=\bar{T}_c)} (x-x_c)\\\nonumber
&&+[(\frac{\partial \bar{Q}}{\partial \bar{T}})]_{(x=x_c, \bar{T}=\bar{T}_c)} (\bar{T}-\bar{T}_c)\\\nonumber
&&+[(\frac{\partial ^2 \bar{Q}}{\partial x^2})]_{(x=x_c, \bar{T}=\bar{T}_c)} (x-x_c)^2\\\nonumber
&&+[(\frac{\partial ^2 \bar{Q}}{\partial \bar{T} \partial x})]_{(x=x_c, \bar{T}=\bar{T}_c)} (x-x_c)(\bar{T}-\bar{T}_c)\\\nonumber
&&+[(\frac{\partial ^3 \bar{Q}}{\partial x^3})]_{(x=x_c, \bar{T}=\bar{T}_c)} (x-x_c)^3\\\nonumber
&&+ higher \ order  \ terms.
\eea
Since the specific heat $C_{\bar{Q}}$ is divergent and $(\frac{\partial \bar{Q}}{\partial \bar{\Phi}})_c=(\frac{\partial ^2 \bar{Q}}{\partial \bar{\Phi}^2})_c=0$ at the critical point, we can rewrite Eq. (\ref{11c}) as follows:
\bea \label{11cc}
\bar{Q}(x)&=&\bar{Q}(x_c)+[(\frac{\partial \bar{Q}}{\partial \bar{T}})]_{(x=x_c, \bar{T}=\bar{T}_c)} (\bar{T}-\bar{T}_c)\\\nonumber
&&+[(\frac{\partial ^2 \bar{Q}}{\partial \bar{T} \partial x})]_{(x=x_c, \bar{T}=\bar{T}_c)} (x-x_c)(\bar{T}-\bar{T}_c)\\\nonumber
&&+[(\frac{\partial ^3 \bar{Q}}{\partial x^3})]_{(x=x_c, \bar{T}=\bar{T}_c)} (x-x_c)^3.
\eea
For the critical exponent $\beta$, we replace Eq. (\ref{11b}) and the differential of Eq. (\ref{11cc}) in Maxwell's equal area law. In this case, we find $\beta=\frac{1}{2}$. In order to obtain the critical exponent $\gamma$ associated with  $k_{\bar{T}}=-\frac{1}{\bar{\Phi}}(\frac{\partial \bar{\Phi}}{\partial \bar{Q}})_{\bar{T}}\propto \epsilon^{-\gamma} $, we differentiate Eqs. (\ref{11b}) and (\ref{11cc}) with respect to $x$ at a constant $\bar{T}$,
 \bea \label{11d}
 k_T = -\frac{\bar{T}_c[(\frac{\partial \bar{\Phi}}{\partial x})_{\bar{Q}}]_{x=x_c}}{\bar{\Phi} [(\frac{\partial ^2 \bar{Q}}{\partial \bar{T} \partial x})]_{(x=x_c, \bar{T}=\bar{T}_c)} } \epsilon ^{-1} \ \Longrightarrow \gamma =1.
\eea
For the critical isotherm $\delta$ described at $\bar{T}=\bar{T}_c$, we can rewrite Eq. (\ref{11cc}) as follows:
\bea \label{11e}
\bar{Q}(x) &=&\bar{Q}(x_c)+\frac{[(\frac{\partial ^3 \bar{Q}}{\partial x^3})]_{(x=x_c, \bar{T}=\bar{T}_c)}}{x_c ^3} \Delta ^3.
\eea
\begin{figure}
  \includegraphics[angle=0,width=13cm,height=8cm]{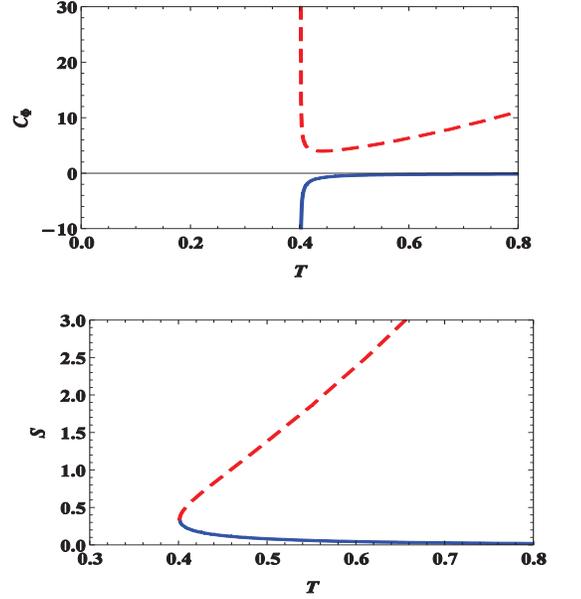}\\
  \caption[] {\it{Specific heat $C_{\Phi}$ and entropy $S$  with respect to
   $T$ for $\Phi=0.4$ and $P=0.3$. }}\label{figure 3}
 \end{figure}
Using Eqs. (\ref{11b}) and (\ref{11e}), we find $\delta =3$. These results demonstrate that the two  methods described above give the same results for the critical exponents if the calculations are performed at the critical point.

\section{ The Critical behavior of charged-AdS black holes }

\noindent In this section, we study the critical behavior of RN-AdS and Gauss-Bonnet-AdS black holes in the ensemble in which pressure $P$ and electric potential $\Phi$ are considered as natural variables. In this case, we define the Gibbs free energy in an extended phase space by
\bea \label{11ee}
G=U+PV-Q \Phi-TS=M-TS-Q \Phi
\eea
where $P=-\frac{\Lambda}{8\pi}=\frac{3}{8 \pi l^2}$ . Also,  mass $M$ is identified with enthalpy rather than with internal energy as long as $P$ is included as a variable in the first law of thermodynamics.
The critical behaviors of Reissner-Nordstrom-AdS and Gauss-Bonnet-AdS black holes in the canonical ensemble and an extended phase space are considered in \cite{Mann1,Mann2,GS}.

\subsection{ The critical behavior of the RN-AdS black hole}

Let us review the thermodynamic quantities of the  RN-AdS black hole. The metric is given by
\bea \label{12}
 ds^2=-f(r) dt^2+\frac{dr^2}{f(r)}+r^2 d\Omega _{2} ^{2},
\eea
\begin{figure}
  \includegraphics[angle=0,width=8cm,height=5cm]{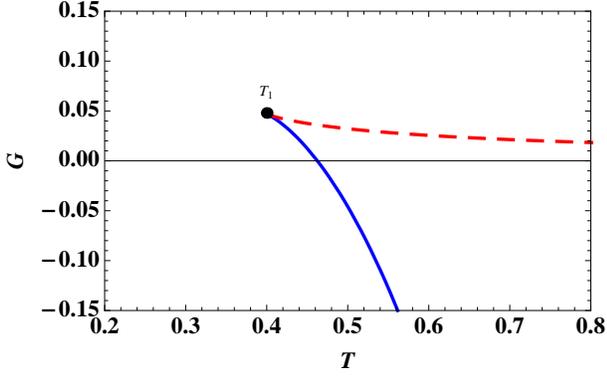}\\
  \caption{\it{Gibbs free energy $G$ with respect to
   $T$ for $\Phi=0.4$ and $P=0.3$. Dashed red and solid blue lines
correspond to positive and negative $C_{\Phi}$, respectively. At $T= T_1$, specific heat is divergent, and we have a second order phase transition. Also, at this point the entropy is continuous and specific heat  is noncontinuous.} }\label{figure 4}
 \end{figure}
where, $f(r)=1-\frac{2M}{r}+\frac{Q^2}{r^2}+\frac{r^2}{l^2}$. We consider the cosmological constant as a thermodynamic pressure $P=-\frac{\Lambda}{8\pi}$ and a conjugate thermodynamic volume as corresponding to $V=\frac{4}{3} \pi r^3_{+}$ \cite{Mann1}. The black hole temperature, entropy, and specific heat are given by
\bea \label{13}
 &&T=\frac{1}{4\pi r_+}(1+8 \pi P r^2_{+}-\Phi ^2)\\
 &&S=\pi r^2 _{+}\\
 &&C_{\Phi}=\frac{1 - \Phi ^2 +8 P \pi r^2_{+}}{2 (4 P - \frac{1 - \Phi ^2 + 8 P \pi r^2_{+}}{4 \pi r^2_{+}})},
\eea

\begin{figure}
\includegraphics[width=12cm]{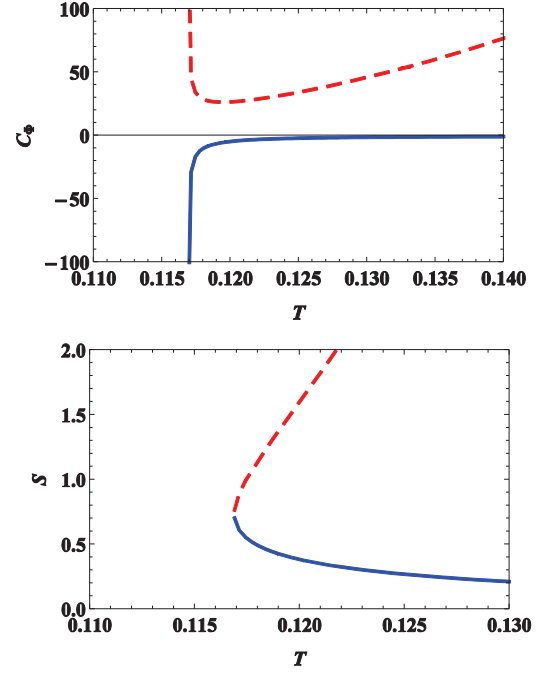}\\
\caption[] {\it{Specific heat $C_{\Phi}$ and entropy $S$  with respect to
$T$ for $d=6$, $\Phi=0.002$, $\alpha=1$, and $P=0.04$. }}\label{figure 5}
\end{figure}
Since $C_{\Phi}$ is divergent at $r_{c_+} =\frac{\sqrt{1 - \Phi^2}}{2 \sqrt{2\pi P}}$, we have a second order phase transition at $r=r_{c_+}$. The Gibbs free energy in the ensemble with pressure $P$ and electric potential $\Phi$ as natural variables is
\bea \label{14}
 G=M-TS-Q \Phi=\frac{1}{4}(r_+ -\frac{8\pi}{3} P r^3 _+ -\Phi ^2 r_+).
\eea
Using Eqs. (\ref{13}) and (\ref{14}), we plot $G$, $C_{\Phi}$ and $S$ with respect to $T$ for fixed values of $\Phi$ and $P$ in Figs. \ref{figure 3} and  \ref{figure 4}. For $T\geq T_1$, there are two wings which are connected at $T_1$. Since the upper wing (solid line) is not the phase with the lowest values of Gibbs free energy of the system and also because it has a negative value for specific heat, the black hole cannot stay in this unstable phase and  falls into the more stable phase with minimum Gibbs free energy and positive specific heat (dashed line) at $T= T_1$.  Specific heat $C_{\Phi}$ is divergent at $T= T_1$ and also flips from positive to negative infinity. Therefore, for the given values of $\Phi$ and $P$, the black hole experiences just a second order phase transition at $T=T_1$. The results are the same for a zero value of $\Phi$, i.e.,  the Schwarzschild-AdS black hole,  the  RN-AdS black hole, in the grand canonical ensemble, and the nonextended phase space \cite{Mann5,B1}.

\subsection{ The critical behavior of the Gauss-Bonnet-AdS black hole (in $d\geq 6$)}

The metric of the  d-dimensional static charged Gauss-Bonnet (GB)-AdS black hole is given by \cite{GS}
\bea \label{15}
ds^2=-f(r)dt^2 +f^{-1} (r) dr^2 +r^2 f_{ij} dx^i dx^j,
\eea
and
\bea \label{16}
f(r)=k+\frac{r^2 }{2 \alpha} (1-\sqrt{1-\frac{4\alpha}{l^2}}\sqrt{1+\frac{m}{r^{d-1}}-\frac{q^2}{r^{2d-4}}}),
\eea
where $\alpha=(d-3)(d-4)\alpha_{GB}$. Mass $M$, temperature $T$,  entropy $S$, and potential $\Phi$ can be described as follows:
\bea \label{17}
&&M=\frac{\Sigma _k}{32 \pi ^2 (d^2 -4d+3) r^{d+5} _{+}}\times ((d-1) Q^2 r^8 _{+}\\\nonumber
&&+ 2 \pi r^{2d} _{+} (d-3) ((d^2-3d+2)(k r^2 _{+} +k^2 \alpha)-2 \Lambda r^4 _{+})),\\
&&T=\frac{1}{8 \pi ^2 (d-2) r^{2d+1} _{+} (2k\alpha +r^2 _{+})}\times(-Q^2 r^8 _{+}\\\nonumber
&&+2\pi r^{2d} _{+}((d-2)k((d-3) r^2 _{+}+(d-5) k \alpha)-2 \Lambda r^4 _{+})),\\
&&S=\frac{\Sigma _k}{4r^{2-d} _{+}} (1+\frac{2k\alpha (d-2)}{(d-4) r^2 _{+}}),\\
&&\Phi =\frac{Q \Sigma _k}{16 \pi ^2 (d-3) r^{d-3} _{+}},
\eea

\noindent where $\Lambda=-8\pi P=-\frac{(d-1)(d-2)}{2l^2}$ and $\Sigma _k$ is the volume of a $(d-2)$-dimensional hypersurface with a constant curvature $(d-2)(d-3)k$.
The specific heat $C_\Phi$ is given by
\bea \label{18}
&&C_\Phi =T(\frac{\partial S}{\partial T})_\Phi\\\nonumber
&&=-\frac{2 \Sigma _k (-2 + d)^2  \pi r^{-3+d} (2 \alpha k + r^2)^3}{\Delta} T,
\eea
where
\bea
\Delta&=& 2 \alpha^2 (-5+d) (-2+d) k^3 + (-3+d) ((-2+d) k \\\nonumber
&-&128 (-3+d)\Phi^2 \pi^3) r^4-\frac{(d-1)(d-2)}{l^2} r^6 + \alpha k r^2\\\nonumber
& (&(-9 + d) (-2 + d) k + 256 (-3 + d)^2 \Phi^2 \pi^3\\\nonumber
& -&\frac{12 (d-1)(d-2)}{2l^2} r^2).
\eea
The specific heat is divergent when $\Delta =0$. Thus, this critical point is given by
\bea \label{19}
&&\Phi _c=\pm \frac{1}{8\sqrt{2} (\pi^{\frac{3}{2}} \sqrt{(3-d)^2 r^2 (-2 \alpha k + r^2)}}\\\nonumber
&&\times (-2 \alpha ^2 (-5+d) (-2+d) k^3 \\\nonumber
&&-\alpha (-9+d) (-2+d) k^2 r^2 -k (6 + (-5+d) d \\\nonumber
&&-\frac{12 \alpha (d-1)(d-2)}{2l^2}) r^4-\frac{(d-1)(d-2)}{l^2} r^6)^{\frac{1}{2}}.
\eea

\begin{figure}
  \includegraphics[width=7cm]{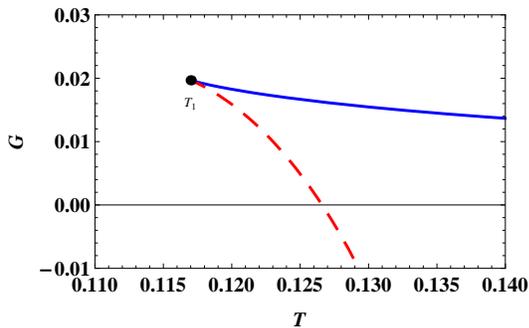}\\
  \caption[] {\it{Gibbs free energy $G$ with respect to
   $T$ for $d=6$, $\Phi=0.002$, $\alpha=1$, and $P=0.04$. Dashed red and solid blue lines
correspond to positive and negative $C_{\Phi}$, respectively. At $T= T_1$, specific heat is divergent and we have a second order phase transition.}}\label{figure 6}
 \end{figure}
 Let us consider the variation of the cosmological constant $\Lambda$ as the thermodynamic pressure $P=\frac{(d-1)(d-2)}{16\pi l^2}$ in an extended phase space  with  pressure and electric potential $\Phi$ as the natural variables. The conjugate thermodynamic volume corresponds to:
\bea
V=(\frac{\partial M}{\partial P})_{S,Q}=\frac{r^{d-1}}{d-1}
\eea
In this case, the Gibbs free energy in the ensemble with pressure $P$ and electric potential $\Phi$ as natural variables is described as
\bea \label{20}
&&G=M-TS-Q\Phi \\\nonumber
&&=\frac{1}{16 \pi (d-4) (d-2) (d-1) (r^2 +2)}\times\\\nonumber
&&[ r^{d-5} (2 (d-2)^2 (d-1) -\frac{2 (d-4)(d-1)(d-2)}{2l^2} r^6\\\nonumber
&&+ (d-2) (d-1) (d-8+256 (d-3) \Phi ^2 \pi ^3) r^2\\\nonumber
&&+( (d-2) (4 + (d-5) d -\frac{12(d-1)(d-2)}{2l^2}\\\nonumber
&&- 128 (d-4) (d-3) (d-1) \Phi ^2 \pi ^3) r^4))],
\eea
where $\Sigma _k$ and $k$ are equal to $1$. Now,  using  Eqs. (\ref{17}), (\ref{18}), and (\ref{20}), we can plot the Gibbs free energy $G$, the specific heat $C_\Phi$, and the entropy $S$ with respect to  temperature $T$ in Figs. \ref{figure 5} and \ref{figure 6}. In the $G-T$ diagram, we have two wings which are connected at $T=T_1$. At phase 1 (solid line), Gibbs free energy has the positive values for all temperatures, so the black hole is unstable in this phase, and it falls into the more stable phase (dashed line) by minimizing the free energy at $T=T_1$. Since the specific heat is divergent at $T=T_1$, the black hole turns into the stable phase (dashed line) through a second order phase transition at  $T=T_1$. These results are similar to the RN-AdS black hole in the extended phase space.\\

\section{Conclusions}
In this paper, the critical behaviors of  static and spherically symmetric hairy black holes were investigated as a solution of massive gravity. The critical exponents of these black holes in the canonical ensemble were calculated. We found them  analogous to the critical behavior and critical exponents of the liquid-gas system in the Van der Waals fluid. In the second part of this paper, we investigated the critical behaviors of Reissner-Nordstrom-AdS and Gauss-Bonnet-AdS black holes in the grand canonical ensemble and the extended phase space by variation of the cosmological constant. In this case,  the black holes were found to experience a second order phase. The results are generalizable to all values of $\Phi$ and $P$; i.e., for $\Phi=0$ (Schwarzschild black hole),  the black hole experiences just a second order phase transition.
%%%%%%%%%%%%%%%%%%%%%%%%%%%%%%%%%%%%%%%%%%%%%%%%%%%%%

%%%%%%%%%%%%%%%%%%%%%%%%%%%%%%%%%%%%%%%%%%%%%%%%%%%%%%%%%%%%%%%%%%%%%%%%%%%%

\end{document}